\documentclass[a4paper]{jpconf}
\usepackage{graphicx}
\usepackage{amssymb}
\usepackage{bm}
\begin{document}

\title{Dynamical alignment of visible and dark sector gauge groups}

\author{Rainer Dick}

\address{Department of Physics and Engineering Physics, 
University of Saskatchewan, 116 Science Place, Saskatoon,
 Canada SK S7N 5E2}

\ead{rainer.dick@usask.ca}

\begin{abstract}
We discuss a dark family of lepton-like particles with their own ``private''
gauge bosons $\bm{X}_\mu$ and $C_\mu$ under a local $SU'(2)\times U'(1)$ symmetry. 
The product of dark and visible gauge 
groups $SU'(2)\times U'(1)\times SU_w(2)\times U_Y(1)$ is broken dynamically to
the diagonal (vector-like) subgroup $SU(2)\times U(1)$ through the coupling of 
two fields $\underline{M}_i$ to the Higgs field and the dark lepton-like particles.
After substituting vacuum expectation values for the fields $\underline{M}_i$, 
the Higgs doublet couples in the standard way to the left-handed $SU'(2)$ 
doublet $\Psi_L$ and right-handed singlets $\psi_{1,R}$, $\psi_{2,R}$,
but not to the extra gauge bosons. This defines a new Higgs portal, where 
the ``dark leptons'' can contribute to the dark matter and interact with Standard 
Model matter through Higgs exchange. It also defines a dark matter model with 
internal interactions.
At low energies, the Standard Model Higgs boson aligns the two
electroweak-type symmetry groups in the visible and dark sectors
and generates the masses in both sectors.
We also identify charge assignments for $SU'(2)\times U'(1)$
in the dark sector which allow for the formation of 
dark atoms as bound states of dark lepton-like particles. 
The simplest single-component dark matter version 
of the model predicts a dark matter mass around 96 GeV, but the 
corresponding nucleon recoil cross section of 
$1.2\times 10^{-44}\,\mathrm{cm}^2$ is ruled out by the xenon based
experiments. However, multi-component models 
or models with a dark $SU'(2)$ doublet mediator
instead of the Higgs portal would still be viable.
\end{abstract}

\section{Introduction}
\label{sec:intro} 

The identification of the dark matter which dominates the large scale structure
in the universe remains an open problem. Higgs exchange had been proposed as an
option for non-gravitational interactions between dark matter and ordinary 
matter \cite{zee,JMcD,bento,cliff,dklm,wells,frank,kusenko}, and Higgs portal models 
with couplings to the scalar product $H^+H$ of the Higgs doublet $H=(H^+,H^0)$ have 
been discussed extensively for bosonic \cite{next1,rd1,yaguna1,yaguna2,MDMvac1,MDMvac2,
MDMvac2b,MDMind4,maxim,batell,djouadi,cline,rdfs1,FPU,rdfs,tom,mariana,V2} 
and fermionic \cite{djouadi,mike,simple,prof2} dark matter, but have meanwhile been 
ruled out in the lower region of the preferred WIMP mass range between about 100 GeV 
and about 1 TeV \cite{pandax,lux2,xenon1t}. So far the traditional Higgs portal  
couplings still remain an option for light dark matter, or for WIMPs heavier 
than 1 TeV. 

In the present paper we suggest studies of another kind of Higgs portal which does 
not involve the scalar product $H^+H$ of the Higgs doublet, but may contribute to 
the masses of dark fermions, although the Higgs field and the dark fermions transform 
under \textit{a priori} separate $SU(2)$ tranformations at high energies.

Mass terms in the dark sector are usually inserted by hand or generated by a 
separate symmetry breaking mechanism. However, it is an intriguing question 
whether $SU(2)$ breaking by the Standard Model (SM) Higgs boson could also generate 
the dark matter masses without violating the Standard Model gauge symmetries, 
and yet be safe from constraints arising through couplings to the electroweak 
gauge bosons. We propose a mechanism to achieve this. The key idea is to have 
an \textit{a priori} separate $SU'(2)\times U'(1)$ gauge symmetry in the dark 
sector with its own gauge bosons $\bm{X}_\mu$ and $C_\mu$. Dynamical breaking
of the dark symmetry, e.g. through scalar fields $\underline{M}_i\equiv\{M_{i,ab}\}$, 
which are charged both under the dark and visible gauge groups, then induces
standard chiral Yukawa couplings of dark left-handed $SU'(2)$ lepton-like doublets 
and right-handed $SU'(2)$ singlets with the Standard Model Higgs doublet,
\begin{equation} \label{eq:HDM1}
\mathcal{L}_{H-DM}=-\frac{\sqrt{2}}{v_h}\left(m_2
\overline{\Psi}_L\cdot H\cdot\psi_{2,R}+
m_1\overline{\Psi}_L\cdot\underline{\epsilon}\cdot 
H^*\cdot\psi_{1,R}\right)+h.c.
\end{equation} 
These Yukawa couplings in turn break the gauge symmetry of the coupled SM+dark matter 
model to $SU_c(3)\times SU_w(2)\times U_Y(1)$, because the Higgs couplings align 
the local $SU(2)\times U(1)$ gauge transformations in the visible and dark sectors. 

Irrespective of whether a variant of the model from Sec. 2 turns out to generate 
the low energy Higgs alignment (\ref{eq:HDM1})
of visible and dark sector gauge groups, or 
whether some other mechanism is at work, the dark gauge groups should also be 
broken up to a possible remnant $U(1)$, in order not to generate too many
dark radiation degrees of freedom. This leaves a remnant
Yukawa potential interaction in the low energy sector if the remnant dark $U(1)$
is broken at low energy, or a dark photon. Dark photons comply with Planck's CMB
constraints on dark radiation if they decouple early enough \cite{weinberg,planck},
and additional relativistic degrees of freedom would also favor a higher 
value of $H_0$, thus reducing the tension between Planck and the 
astronomical measurements of the Hubble constant. 
The dark leptons, on the other hand, acquire mass terms through their SM-like 
Yukawa couplings to the Higgs (and possibly also due to internal symmetry
breaking in the dark sector) and can constitute the dark 
matter, assuming that leptogenesis also had an analog
in the dark sector. 
Neutrality under the internal gauge interactions of the dark sector
then implies that the dark matter will consist of dark atoms.
The resulting dark matter model is therefore similar to the dark atom
models, see \cite{kaplan,sigurdson,cline2,petraki} and references there.
The primary new idea is a dynamically generated Higgs alignment between 
visible and dark gauge groups.

Breaking of dark gauge symmetries occurs in particular if
the Higgs boson couples both to the visible and dark
electroweak type gauge bosons,
\begin{eqnarray} \nonumber
D_\mu H(x)&=&\partial_\mu H(x)
-ig_w\bm{W}_\mu(x)\cdot\frac{\bm{\sigma}}{2}H(x)
-i\frac{g_Y}{2}B_\mu(x)H(x)
\\ \label{eq:DHfull}
&&-\,iq_2\bm{X}_\mu(x)\cdot\frac{\bm{\sigma}}{2}H(x)
-i\frac{q_1}{2}Y'_h C_\mu(x)H(x).
\end{eqnarray}
Combined with Eq.~(\ref{eq:HDM1}), this defines a class of 
renormalizable Higgs portal models for dark matter with
 $SU_w(2)\times U_y(1)\times SU'(2)\times U'(1)$ gauge symmetry
and spontaneously broken $SU(2)$ factors. This model would not be
ruled out (yet) through bounds on the invisible Higgs decay
width if all massive particles in the dark sector are heavier 
than $m_h/2$, and if all dark particles with masses below $m_h/2$
have small masses, $m\ll m_h/2$.

However, as outlined above, in the present investigation we
will focus on the case that the Higgs boson does not directly
couple to the $SU'(2)\times U'(1)$ gauge bosons, but is
only charged under the Standard Model gauge group,
 \begin{equation} \label{eq:DHSM}
D_\mu H(x)=\partial_\mu H(x)
-ig_w\bm{W}_\mu(x)\cdot\frac{\bm{\sigma}}{2}H(x)
-i\frac{g_Y}{2}B_\mu(x)H(x).
\end{equation}
Section \ref{sec:origin} provides a dynamical toy model which generates
the Higgs portal coupling (\ref{eq:HDM1}), although the covariant
derivative on the Higgs field is only given by (\ref{eq:DHSM}).
Section \ref{sec:model} then provides a more fulsome discussion of 
the low energy formulation of the theory, and the formation of dark 
atoms in this theory is discussed in Sec.~\ref{sec:atoms}. 
Annihilation cross sections and the constraints from thermal dark 
matter creation on the simplest version of dark matter in 
this framework are discussed in Section \ref{sec:thermal}. 
Section \ref{sec:conc} summarizes our conclusions.

\section{A dynamical mechanism for alignment of visible
 and dark gauge symmetries}
\label{sec:origin}

A dynamical mechanism to generate the coupling (\ref{eq:HDM1})
with a Higgs field which is not charged under the dark gauge groups,
cf.~(\ref{eq:DHSM}), can be constructed with 
scalar fields $\underline{M}_i\equiv\{M_{i,ab}\}$ which are charged
both under the dark gauge group and the Standard Model gauge group.
The first index $a$ refers to the fundamental representation
of $SU'(2)$ from the left, while the second index $b$ refers to $SU_w(2)$ 
acting through the adjoint $SU_w(2)$-matrices from the right. The $U'(1)$ 
charges are given in terms of the dark fermion charges 
by $Y'^{(M)}_{i}=Y'_L-Y'_{i,R}$, and the $U_Y(1)$ charges 
are $Y^{(M)}_{i}=-Y_h=-1$. The covariant derivatives 
of the $\underline{M}_i$-fields are therefore
\begin{eqnarray}\nonumber
D_\mu\underline{M}_i&=&\partial_\mu\underline{M}_i-iq_2\bm{X}_\mu\cdot\frac{\bm{\sigma}}{2}
\cdot\underline{M}_i-i\frac{q_1}{2}\left(Y'_L-Y'_{i,R}\right)C_\mu\underline{M}_i
\\ \label{eq:DMi}
&&+\,ig_W\underline{M}_i\cdot\frac{\bm{\sigma}}{2}\cdot\bm{W}_\mu
+i\frac{g_Y}{2}Y_hB_\mu\underline{M}_i.
\end{eqnarray}
The following coupling term between the visible and the dark sector is then
invariant under the full gauge group $SU'(2)\times U'(1)\times SU_w(2)\times U_Y(1)$,
\begin{equation} \label{eq:MHDM1}
\mathcal{L}_{MH-DM}=-\frac{\sqrt{2}}{v_h}\left(
\overline{\Psi}_L\cdot\underline{M}_2\cdot H\cdot\psi_{2,R}
+
\overline{\Psi}_L\cdot\underline{M}_1\cdot\underline{\epsilon}\cdot 
H^*\cdot\psi_{1,R}\right)+h.c.
\end{equation} 
This yields the new Higgs portal coupling (\ref{eq:HDM1})
in the low energy sector through the potential
\begin{equation}\label{eq:potMi}
V(\underline{M}_1,\underline{M}_2)=\frac{1}{4}\sum_{i=1}^2
\lambda_i\left[\mathrm{Tr}\left(\underline{M}_i\cdot\underline{M}_i^+\right)
-2\mathrm{Det}\underline{M}_i\right]^2.
\end{equation}
A $2\times 2$ matrix $\underline{M}_i$ satisfies the ground state condition
\begin{equation}\label{eq:groundMi1}
\mathrm{Tr}\left(\underline{M}_i\cdot\underline{M}_i^+\right)
=2\mathrm{Det}\underline{M}_i
\end{equation}
if and only if the matrix is proportional to the unit matrix,
\begin{equation}\label{eq:groundMi2}
\underline{M}_i=m_i\underline{1},
\end{equation}
up to an additional possible unitary factor $\underline{V}_i$. The 
parameter $m_i$ in Eq. (\ref{eq:groundMi2}) can be chosen to 
satisfy $m_i\ge 0$. These results can easily be proved using the polar 
decomposition $\underline{M}_i=\underline{H}_i\cdot\underline{V}_i$
of the matrix $\underline{M}_i$ into a positive semidefinite 
hermitian factor $\underline{H}_i$ and a unitary factor $\underline{V}_i$. 

This simple dynamical model is only a toy 
model for the demonstration that the new Higgs portal (\ref{eq:HDM1}) can arise
dynamically, if we are primarily interested in the traditional WIMP mass range, 
which will be the focus of the remainder of this paper. Any actual phenomenological
implementation of the dynamical model (\ref{eq:DMi}-\ref{eq:potMi}) could only work
if dark matter is extremely light.

\section{Dark matter masses from the Standard Model Higgs boson}
\label{sec:model} 

 For all we know, the Higgs field is the only dynamical field (besides gravity)
which couples to all massive fields, and indeed generates the masses of
those fields through Yukawa couplings. It is therefore compelling to assume that
the Higgs field also generates the masses in the dark sector. The Yukawa 
couplings (\ref{eq:HDM1}) to fermionic dark matter $\Psi_L=(\psi_{1,L},\psi_{2,L})$,
 $\psi_{1,R}$ and $\psi_{2,R}$ provide a way to achieve this, and the mechanism 
outlined in Sec. \ref{sec:origin} provides a way to generate these
couplings in the low energy sector of a renormalizable gauge theory.

We assume a single generation of dark lepton-like particles in (\ref{eq:HDM1}),
but the generalization to more generations is straightforward through promotion 
of the masses $m_1$ and $m_2$ to mixing matrices.

\textit{A priori} the dark gauge group $SU'(2)\times U'(1)$ acting on 
the fields $\Psi_L$ and $\psi_{i,R}$ and the electroweak gauge group are 
different symmetries, with $SU'(2)\times U'(1)$ acting only in the dark 
sector while the electroweak symmetry only acts in the visible sector.
However, the Yukawa couplings (\ref{eq:HDM1}) of the Higgs 
doublet \textit{align} the transformations in both symmetry groups at low energies, 
thus breaking the direct product of symmetry groups to its diagonal component,
\begin{equation} \label{eq:break1}
SU'(2)\times U'(1)\times SU_w(2)\times U_Y(1)
\to SU'(2)\times U'(1)=SU_w(2)\times U_Y(1).
\end{equation}
The corresponding dark hypercharge ($U'(1)$) assignments have to satisfy
\[
Y'_L-Y'_{1,R}=-Y_h=-1,\quad
Y'_L-Y'_{2,R}=Y_h=1,
\]
where $Y_h=1$ is the weak hypercharge of the Higgs doublet.

The gauge symmetry $SU_w(2)\times U_Y(1)$ is therefore implemented for 
low energy in the visible and dark sectors through the $SU(2)$ 
transformations $U(x)=\exp[i\bm{\varphi}(x)\cdot\bm{\sigma}/2]$
and the $U(1)$ transformations $\exp[i\alpha(x)Y/2]$,
$\exp[i\alpha(x)Y'/2]$. The action of the SM fields is as usual, 
\[
H'(x)=\exp[i\alpha(x)/2]U(x)\cdot H(x),
\]
\[
\left(\begin{array}{c}
\nu'_L(x) \\ e'_L(x)\\
\end{array}\right)=\exp[-i\alpha(x)/2]U(x)\cdot 
\left(\begin{array}{c}
\nu_L(x) \\ e_L(x)\\
\end{array}\right),
\]
\[
\nu'_R(x)=\nu_R(x),\quad e'_R(x)=\exp[-i\alpha(x)]e_R(x),\ldots
\]
\begin{equation}\label{eq:Wprime}
\bm{W}'_\mu(x)\cdot\bm{\sigma}
=U(x)\cdot (\bm{W}_\mu(x)\cdot\bm{\sigma})\cdot U^+(x)
+\frac{2i}{g_w}U(x)\cdot\partial_\mu U^+(x),
\end{equation}
\[
B'_\mu(x)=B_\mu(x)+\frac{1}{g_Y}\partial_\mu\alpha(x),
\]
and the corresponding action in the dark sector is
\begin{equation} \label{eq:Psiprime}
\Psi'_L(x)\equiv
\left(\begin{array}{c}
\psi'_{1,L}(x) \\ \psi'_{2,L}(x)\\
\end{array}\right)
=\exp[i Y'_L\alpha(x)/2]U(x)\cdot 
\left(\begin{array}{c}
\psi_{1,L}(x) \\ \psi_{2,L}(x)\\
\end{array}\right),
\end{equation}
\[
\psi'_{1,R}(x)=\exp[i (Y'_L+1)\alpha(x)/2]\psi_{1,R}(x),
\quad
\psi'_{2,R}(x)=\exp[i (Y'_L-1)\alpha(x)/2]\psi_{2,R}(x),
\]
\begin{equation} \label{eq:Xprime}
\bm{X}'_\mu(x)\cdot\bm{\sigma}
=U(x)\cdot (\bm{X}_\mu(x)\cdot\bm{\sigma})\cdot U^+(x)
+\frac{2i}{q_2}U(x)\cdot\partial_\mu U^+(x),
\end{equation}
\[
C'_\mu(x)=C_\mu(x)+\frac{1}{q_1}\partial_\mu\alpha(x).
\]
Here $q_2$ and $q_1$ are the gauge couplings of the dark $SU'(2)\times U'(1)$ 
symmetry group, and the covariant derivatives on the dark matter fields are
\begin{equation} \label{eq:DPsiL}
D_\mu\Psi_L(x)=\partial_\mu\Psi_L(x)
-iq_2\bm{X}_\mu(x)\cdot\frac{\bm{\sigma}}{2}\Psi_L(x)
-i\frac{q_1}{2}Y'_L C_\mu(x)\Psi_L(x),
\end{equation}
\begin{equation}\label{eq:DPsiR}
D_\mu\psi_{i,R}(x)
=\left(\partial_\mu-i\frac{q_1}{2}Y'_{i,R} C_\mu(x)\right)\psi_{i,R}(x),
\end{equation}
\begin{equation}\label{eq:YiR}
Y'_{i,R}=Y'_L-(-)^i.
\end{equation}
The low energy sector of the theory outlined in Sec. \ref{sec:origin} therefore
looks like the Standard Model augmented with a fourth lepton-like family with
its own ``private gauge bosons'' $\bm{X}_\mu$ and $C_\mu$. Since we assume $m_i>m_h/2$,
this ``fourth family'' does not contribute to $Z$ nor Higgs decays, nor does it
produce a fourth low mass thermalized particle.

The reader will certainly have noticed that the assumed preference for 
left-chirality
also in the dark sector is not determined by any experimental observation
and therefore ambiguous with our current knowledge.
Instead, we could just as well use right-handed $SU'(2)$ doublets and left-handed
singlets in the dark sector. This would be compelling from an enhanced
symmetry point of view. Opposite chirality in the dark sector
could restore $CP$ and time-reversal symmetry in particle physics through
mappings between visible sector and dark sector fields. $CP$ and time
reversal symmetry might then only be hidden from us through the very weak 
Higgs-coupling between the visible sector and the dark sector.
The model would then provide a Higgs portal realization of the old idea 
of a $CP$ mirror of the Standard Model \cite{hodges}.
However, currently we can constrain the dark sector
only through gravitational effects and direct and indirect search limits.
We cannot currently distinguish between a $V-A$ or $V+A$ preference in the 
dark sector, and at this stage a $V-A$ formulation is as good as 
a $V+A$ formulation.

There are obviously infinitely many possible ramifications of this model
by also including corresponding families of dark quarks with dark symmetry
group $SU'_c(N)$ and corresponding gauge bosons $F_\mu^a$. The Yukawa couplings
of the Higgs boson to the dark quark-like multiplets would obviously not align
the $SU'_c(N)$ with the Standard Model $SU_c(3)$. In such a model
it would be natural to expect the dark matter to consist of dark 
atoms formed from dark leptons and dark nucleons, with the atoms formed due 
to the $SU'(2)\times U'(1)$ interactions in the dark sector.

However, we will focus on the simplest model of private gauge bosons with
one dark family of lepton-like particles and their private $SU(2)\times U(1)$ 
gauge bosons (\ref{eq:HDM1},\ref{eq:DPsiL},\ref{eq:DPsiR},\ref{eq:YiR}).
This is practically equivalent to a dark multi-family model with one family 
being lighter than the other families, thus being the relevant family for 
dark matter. Because we do not need to exclude the presence of heavier
lepton-like or quark-like $SU'(2)\times U'(1)$ doublets, we also do not
have to worry about anomaly cancellation within the lightest lepton-like
dark family. However, we will later find that the requirement for dark 
matter in the single lepton-like dark family implies $Y'_L=0$, such that 
anomaly cancellation is actually fulfilled
already with the single dark doublet $\Psi_L$. 

Without dark nucleons from a dark quark sector, 
Coulomb-type repulsion between the dark leptons could
counteract the pull of gravity. Generically this would seem to imply
that the $SU'(2)\times U'(1)$ couplings $q_2$ and $q_1$ should only be of 
gravitational strength for not actually preventing dark halo formation. 
However, we can choose the $SU'(2)\times U'(1)$ charges of the dark leptons 
in a way which allows for the formation of $SU'(2)\times U'(1)$ neutral dark 
atoms as bound states of the dark leptons, thus alleviating the
contraints on the magnitudes of $q_1$ and $q_2$.

\section{Dark atoms}
\label{sec:atoms} 

Recall that the Coulomb potential for electrons and protons
in quantum optics (expressed in terms of Schr\"odinger
picture operators)
\begin{eqnarray} \nonumber
H_C&=&\!\int\!d^3\bm{x}\int\!d^3\bm{x}'\,\frac{\alpha_S}{2|\bm{x}-\bm{x}'|}
\!\left[\psi^+_e(\bm{x})\psi^+_e(\bm{x}')\psi_e(\bm{x}')\psi_e(\bm{x})
\right.
\\ \label{eq:hcqo}
&&
+\left.\psi^+_p(\bm{x})\psi^+_p(\bm{x}')\psi_p(\bm{x}')\psi_p(\bm{x})
-2\psi^+_e(\bm{x})\psi^+_p(\bm{x}')\psi_p(\bm{x}')\psi_e(\bm{x})
\right]
\end{eqnarray}
arises from 
\begin{equation}\label{eq:C1}
\partial_\mu F^{\mu0}=e\left(\psi^+_e\psi_e-\psi^+_p\psi_p\right)
\end{equation}
in Coulomb gauge, and substitution of the solution 
\[
\bm{E}_\|(\bm{x})=\bm{\nabla}\int\!d^3\bm{x}'
\frac{e}{4\pi|\bm{x}-\bm{x}'|}
\left(\psi^+_e(\bm{x}')\psi_e(\bm{x}')-\psi^+_p(\bm{x}')\psi_p(\bm{x}')\right)
\]
of Eq.~(\ref{eq:C1}) into the contribution from $\bm{E}_\|$
to the electromagnetic field Hamiltonian
\begin{equation} \label{eq:hcqo2}
H_{EM}=\frac{1}{2}\int\!d^3\bm{x}\left(\bm{E}^2_\|(\bm{x})
+\bm{E}^2_\perp(\bm{x})+\bm{B}^2(\bm{x})\right)
=H_C+\frac{1}{2}\int\!d^3\bm{x}\left(
\bm{E}^2_\perp(\bm{x})+\bm{B}^2(\bm{x})\right),
\end{equation}
\[
\bm{E}_\perp(\bm{x})=-\partial\bm{A}(\bm{x},t)/\partial t\Big|_{t=0},\quad
\bm{B}(\bm{x})=\bm{\nabla}\times\bm{A}(\bm{x}).
\]

Substitution of the $\bm{k}$-space expansions of the electron and 
proton operators and normal-ordering then yields the usual
electron-electron, electron-(anti-proton), proton-proton, 
positron-proton etc.~repulsion terms,
and the electron-positron, electron-proton etc.~attraction terms.

A corresponding analysis helps to identify any attractive particle-particle 
combinations in the dark sector, although for non-abelian dark interactions 
the Coulomb kernel will only yield the perturbative short-distance part of the 
actual potential if the interaction is asymptotically free, or the Coulomb 
kernel needs to be replaced by a Yukawa kernel for massive gauge fields.

In the non-relativistic limit, the equations
(with $\psi^+_i\psi_i\equiv\psi^+_{i,L}\psi_{i,L}+\psi^+_{i,R}\psi_{i,R}$)
\[
\partial_\mu C^{\mu 0}=-\varrho_C=-\frac{q_1}{2}\Big[Y'_L
\left(\psi^+_1\psi_1+\psi^+_2\psi_2\right)
+\psi^+_{1,R}\psi_{1,R}-\psi^+_{2,R}\psi_{2,R}\Big],
\]
\[
D_\mu X_a^{\mu 0}=-\varrho_a=-\frac{q_2}{2}\Psi^+_L\cdot\sigma_a\cdot\Psi_L,
\]
yield the dark sector Coulomb operator 
\begin{equation}\label{eq:Cd1}
H_{dC}=\int\!d^3\bm{x}\int\!d^3\bm{x}'\,\frac{\varrho_d^2(\bm{x},\bm{x}')
}{8\pi|\bm{x}-\bm{x}'|}
\end{equation}
with the charged density-density correlation operator
\begin{eqnarray} \nonumber
\varrho_d^2(\bm{x},\bm{x}')&=&\varrho_C^2(\bm{x},\bm{x}')
+\sum_{a=1}^3\varrho_a^2(\bm{x},\bm{x}')
\\ \nonumber
&=&\frac{q_2^2}{4}
\left[
\psi^+_{1,L}(\bm{x})\psi^+_{1,L}(\bm{x}')\psi_{1,L}(\bm{x}')\psi_{1,L}(\bm{x})
+\psi^+_{2,L}(\bm{x})\psi^+_{2,L}(\bm{x}')\psi_{2,L}(\bm{x}')\psi_{2,L}(\bm{x})
\right.
\\ \nonumber
&&
-\left.2\psi^+_{1,L}(\bm{x})\psi^+_{2,L}(\bm{x}')\psi_{2,L}(\bm{x}')\psi_{1,L}(\bm{x})
+4\psi^+_{1,L}(\bm{x})\psi^+_{2,L}(\bm{x}')\psi_{1,L}(\bm{x}')\psi_{2,L}(\bm{x})
\right]
\\ \nonumber
&&+\,\frac{q_1^2}{4}Y'^2_L
\left[
\psi^+_{1,L}(\bm{x})\psi^+_{1,L}(\bm{x}')\psi_{1,L}(\bm{x}')\psi_{1,L}(\bm{x})
+\psi^+_{2,L}(\bm{x})\psi^+_{2,L}(\bm{x}')\psi_{2,L}(\bm{x}')\psi_{2,L}(\bm{x})
\right.
\\ \nonumber
&&
+\left.2\psi^+_{1,L}(\bm{x})\psi^+_{2,L}(\bm{x}')\psi_{2,L}(\bm{x}')\psi_{1,L}(\bm{x})
\right]
\\ \nonumber
&&
+\,\frac{q_1^2}{4}
\left[
(Y'_L+1)^2\psi^+_{1,R}(\bm{x})\psi^+_{1,R}(\bm{x}')\psi_{1,R}(\bm{x}')\psi_{1,R}(\bm{x})
\right.
\\ \nonumber
&&
+\,(Y'_L-1)^2\psi^+_{2,R}(\bm{x})\psi^+_{2,R}(\bm{x}')\psi_{2,R}(\bm{x}')\psi_{2,R}(\bm{x})
\\ \nonumber
&&
+\left. 2(Y'^2_L-1)\psi^+_{1,R}(\bm{x})\psi^+_{2,R}(\bm{x}')\psi_{2,R}(\bm{x}')\psi_{1,R}(\bm{x})
\right]
\\ \nonumber
&&
+\,\frac{q_1^2}{2}Y'_L(Y'_L+1)
\left[
\psi^+_{1,L}(\bm{x})\psi^+_{1,R}(\bm{x}')\psi_{1,R}(\bm{x}')\psi_{1,L}(\bm{x})
\right.
\\ \nonumber
&&
+\left.\psi^+_{2,L}(\bm{x})\psi^+_{1,R}(\bm{x}')\psi_{1,R}(\bm{x}')\psi_{2,L}(\bm{x})
\right]
\\ \nonumber
&&+\,\frac{q_1^2}{2}Y'_L(Y'_L-1)
\left[
\psi^+_{1,L}(\bm{x})\psi^+_{2,R}(\bm{x}')\psi_{2,R}(\bm{x}')\psi_{1,L}(\bm{x})
\right.
\\ \label{eq:dC2}
&&
+\left.\psi^+_{2,L}(\bm{x})\psi^+_{2,R}(\bm{x}')\psi_{2,R}(\bm{x}')\psi_{2,L}(\bm{x})
\right].
\end{eqnarray}

Most of the terms in Eqs.~(\ref{eq:Cd1},\ref{eq:dC2}) are repulsive between
pairs of particles and attractive between particles and anti-particles in the
dark sector. The attractive channels between particles and anti-particles 
allow for the formation of dark $SU'(2)$ mesons which will decay fast through 
the Higgs portal (\ref{eq:HDM1}) and accelerate annihilation of any remnant
dark anti-leptons. We are therefore interested in attractive terms between 
pairs of particles for the formation of dark atoms.

The terms in Eq.~(\ref{eq:dC2}) with coupling constants $q_1^2Y'_L(Y'_L\pm 1)$
yield attractive interactions in the two-particle 
states $\psi^+_{1,L}(\bm{x})\psi^+_{1,R}(\bm{x}')\bm{|}0\bm{\rangle}$
and $\psi^+_{2,L}(\bm{x})\psi^+_{1,R}(\bm{x}')\bm{|}0\bm{\rangle}$
if $Y'_L(Y'_L+1)<0$,
and attractive interactions 
in the two-particle 
states $\psi^+_{1,L}(\bm{x})\psi^+_{2,R}(\bm{x}')\bm{|}0\bm{\rangle}$
and $\psi^+_{2,L}(\bm{x})\psi^+_{2,R}(\bm{x}')\bm{|}0\bm{\rangle}$
if $Y'_L(Y'_L-1)<0$. However, the resulting bound states
would have residual charges $q_2/2$ under $SU'(2)$
and $(2Y'_L\pm 1)q_1/2$ under $U'(1)$, respectively. This would yield 
repulsive Coulomb-type atom-atom interactions and again prevent dark
halo collapse unless $q_1$ and $q_2$ would be tuned below gravitational
strength.

However, the Coulomb term for the dark two-lepton 
states $\psi^+_{1,R}(\bm{x})\psi^+_{2,R}(\bm{x}')\bm{|}0\bm{\rangle}$,
\begin{equation} \label{eq:hrr1}
H_{1R,2R}=\int\!d^3\bm{x}\int\!d^3\bm{x}'\,\frac{
q_1^2(Y'^2_L-1)}{16\pi|\bm{x}-\bm{x}'|}
\psi^+_{1,R}(\bm{x})\psi^+_{2,R}(\bm{x}')
\psi_{2,R}(\bm{x}')\psi_{1,R}(\bm{x})
\end{equation}
is attractive if $Y'^2_L<1$. It has no $SU'(2)$ charge, but a 
resulting atomic $U'(1)$ charge $q_1Y'_L$. Setting $Y'_L=0$ 
therefore yields uncharged dark atoms consisting of the two 
right-handed dark leptons. This implies overall charge
neutrality of the universe also under the dark gauge groups, in the same
way as anomaly cancellation in the Standard Model ensures charge neutrality
in the visible sector. At temperatures above 1 TeV there are equal abundances
of particle species, and the vanishing sum of visible $U_Y(1)$ hypercharges
over SM particle species and dark $U'(1)$ hypercharges over dark sector
species ensures overall charge neutralities in both sectors.

The potential in the dark sector therefore splits
into the left-handed and right-handed parts, $H_{dC}=H_{LL}+H_{RR}$,
with fine structure constants $\alpha_L=q_2^2/16\pi$
and $\alpha_R=q_1^2/16\pi$, and the only attractive particle-particle
channel for neutral atoms consisting of the two right-handed
dark leptons, which are bound due to the $U'(1)$ interaction. 
The states of the dark atoms are therefore for 
separation $\bm{r}=\bm{x}_1-\bm{x}_2$
of the two dark leptons and total atomic momentum $\bm{K}$
given in the standard way by direct transcription of the
corresponding results of non-relativistic QFT,
\begin{eqnarray}\nonumber
\bm{|}\Psi_{n,\ell,m_\ell;\bm{K}}(t)\bm{\rangle}
&=&\int\!d^3\bm{x}_1\int\!d^3\bm{x}_2\,
\psi^+_{1,R}(\bm{x}_1)\psi^+_{2,R}(\bm{x}_2)\bm{|}0\bm{\rangle}
\Psi_{n,\ell,m_\ell}(\bm{x}_1-\bm{x}_2)(2\pi)^{-3/2}
\\ \label{eq:da1}
&&\times
\exp[i\bm{K}\cdot(m_1\bm{x}_1+m_2\bm{x}_2)/M]
\exp[-iE(\bm{K},n)t],
\end{eqnarray}
with hydrogen type wave functions $\Psi_{n,\ell,m_\ell}(\bm{r})$
 for coupling constant $\alpha_R/4$ and reduced 
mass $m_{12}=m_1 m_2/M$, $M=m_1+m_2$.
The energy eigenvalues in the non-relativistic regime are
\[
E(\bm{K},n)=\frac{\bm{K}^2}{2M}-\frac{1}{32n^2}\alpha_R^2 m_{12}.
\]
Spin singlet or triplet factors were apparently omitted, since
we are not interested in the fine structure of the dark atoms.
The chiral projectors in the potential (\ref{eq:hrr1}) reduce
the effective dark sector fine structure constant to $\alpha_R/4$.
This is explained in detail in the Appendix.

We also assume $\alpha_R<\alpha_S$ since dark $U(1)$
interactions must be weaker than electromagnetism.
The separation of gas and dark matter in bullet-type clusters
warrants this conclusion \cite{bullet1,bullet2}. 
The fact that dark matter halos are much more extended and form less 
concentrated substructures than baryonic matter also tells us that
dark matter cannot cool down as efficiently as baryons.

\section{Dark matter annihilation and thermal creation}
\label{sec:thermal} 

We are interested in the non-relativistic thermal freeze-out of WIMP 
scale dark leptons. The annihilation of a heavy dark lepton species 
is then dominated by branching ratios into Standard Model particles 
through the Higgs portal (\ref{eq:HDM1}).
The $t$ and $u$ channel annihilations $\psi_{i}\overline{\psi}_{i}\to CC$
and $\psi_{i}\overline{\psi}_{i}\to\bm{X}\bm{X}$ are suppressed 
with $\alpha_R^2\bm{p}^2/m_i^2$ due to the chirality factors in the 
amplitudes, which come from the chirality factors $(1\pm\gamma_5)/2$ in the 
vertices $q_1Y'_{i,R}\overline{\psi}_i\gamma^\mu C_\mu(1+\gamma_5)\psi_i/4$
and 
$q_2\overline{\psi}_i\gamma^\mu\bm{X}_\mu\cdot\bm{\sigma}(1-\gamma_5)\psi_i/4$
 from Eqs.~(\ref{eq:DPsiL},\ref{eq:DPsiR}).

The leading order cross sections into the Standard Model states are
(with $VV=W^+ W^-$ or $VV=ZZ$ and $\delta_W=1$ for annihilation 
into $W^+ W^-$ or $\delta_W=0$ otherwise)
\begin{equation} \label{eq:dd2VV}
\sigma_{\psi_{i}\overline{\psi}_{i}\to VV}(s)
=\frac{1+\delta_W}{64\pi}\frac{m_i^2}{v_h^4 s}
\sqrt{s-4m_V^2}\sqrt{s-4m_i^2}
\frac{(s-2m_V^2)^2+8m_V^4}{(s-m_h^2)^2+m_h^2\Gamma_h^2},
\end{equation}
\begin{equation}\label{eq:dd2ff}
\sigma_{\psi_{i}\overline{\psi}_{i}\to f\overline{f}}(s)
=N_c\frac{m_i^2 m_f^2}{16\pi v_h^4 s}(s-4m_f^2)^{3/2}
\frac{(s-4m_i^2)^{1/2}}{(s-m_h^2)^2+m_h^2\Gamma_h^2},
\end{equation}
\begin{equation}\label{eq:dd2hh}
\sigma_{\psi_{i}\overline{\psi}_{i}\to hh}(s)
=\frac{9m_i^2 m_h^4}{64\pi v_h^4 s}\frac{\sqrt{s-4m_h^2}\sqrt{s-4m_i^2}}{
(s-m_h^2)^2+m_h^2\Gamma_h^2}.
\end{equation}

The total annihilation cross section increases with mass $m_i$ for
masses above 80 GeV, and therefore the heavier dark lepton species 
will determine both the mass $M$ of the dark atoms and the freeze 
out temperature. We will assume $m_1\lesssim 0.01 m_2$ 
and therefore $M\simeq m_2$. 

 \begin{figure}
\begin{center}
 \scalebox{1}{\includegraphics{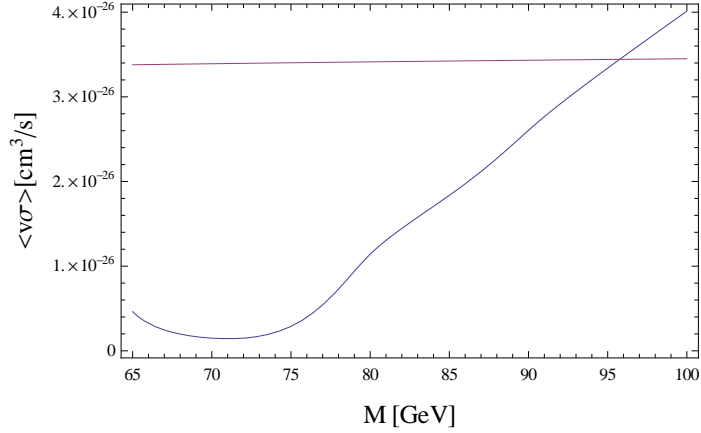}}
\end{center}
 \caption{\label{fig:1} The thermally averaged annihilation cross section
of the dark leptons (in blue) versus the required cross section for thermal
dark matter creation (the logarithmically varying line in purple).}
 \end{figure}

The requirement of thermal freeze out then determines $M\simeq 96$ GeV,
see Fig.~\ref{fig:1}, where the thermally averaged \cite{gondolo} 
annihilation cross for
a particle with mass $m_2\simeq M$ is compared to the required value from
thermal dark matter creation. The logarithmically varying required value 
of $\langle v\sigma\rangle$ for thermal creation 
varies from $3.38\times 10^{-26}\,\mathrm{cm}^3/\mathrm{s}$ for $M=65$ GeV
to $3.45\times 10^{-26}\,\mathrm{cm}^3/\mathrm{s}$ for $M=100$ GeV.

A low mass value $m_1\lesssim 1$ GeV implies an invisible Higgs decay
width which is well below the current limits \cite{atlas,cms},
$\Gamma_{h\to\psi_1\overline{\psi}_1}\lesssim 82\,\mathrm{keV}\simeq 0.018\Gamma_{h\to\mathrm{SM}}$.

This dark matter model is even more predictive than the standard
Higgs portal dark matter models because the coupling to the Higgs field
is already determined in terms of the mass, $g=m_2/v_h\simeq M/v_h$.
The requirement of thermal dark matter creation therefore does not yield
a parametrization $g(M)$ of the Higgs portal coupling as a function of
the dark matter mass, but determines $M$.
However, the corresponding nucleon recoil cross section 
\begin{equation}\label{eq:recoil}
\sigma_{DN}=\frac{g^2}{4\pi}\frac{m_N^2 M^4}{m_h^4 v_h^2(M+m_N)^2}
\end{equation}
is about $1.2\times 10^{-44}\,\mathrm{cm}^2$ for $M=96$ GeV, $m_N=930.6$ 
MeV (the weighted average of the nucleon masses in long lived xenon
isotopes), and the SVZ value $gv_h=210$ MeV \cite{SVZ} for the 
effective Higgs-nucleon coupling. This is in conflict with
the exclusion limits from the xenon based direct 
search experiments \cite{xenon100,lux,pandax,lux2,xenon1t}.

\section{Conclusions}
\label{sec:conc} 

Alignment of gauge symmetries in the visible and dark sectors through
the new Higgs portal is an interesting new tool for dark matter model
building. It can arise as a consequence of dynamical symmetry breaking
in gauge theories, and it opens a door to fermionic Higgs portal 
models without the need of higher mass-dimension effective vertices.

Apparently, the construction presented here opens the Higgs portal
into much more complicated and rich dark sectors, even with the
possibility of $CP$ and time-reversal reciprocity between the
visible and dark sectors, which would be broken through the
different mass spectra in the two sectors. Furthermore, the
construction also generalizes to alignment of gauge groups
through other fields. Every field which transforms in a faithful
representation of a symmetry group $G$ can align different
copies $G_1,G_2,\ldots$ of the symmetry group, each with their own 
gauge bosons $A^a_{i,\mu}$ and coupling constants $g_i$, through 
Yukawa couplings to fields transforming under the different 
symmetries $G_i$. In particular, it is conceivable that the dark
leptons may not couple to the Standard Model through the Higgs 
portal, but through a dark scalar $SU'(2)$ doublet $H'$, which 
couples to the dark sector gauge bosons $\bm{X}_\mu$ and $C_\mu$.
This would also align the electroweak-type gauge symmetries
in the dark and visible sectors through
the Yukawa couplings of the scalar $SU'(2)$ doublet.
However, it would not be constrained by the current limits from
the direct search experiments, since the Yukawa couplings of $H'$ 
in the visible sector must be weaker than the Higgs couplings, 
thus also implying a weaker effective $H'$-nucleon coupling.

\ack
This work was supported in part by the Natural Sciences and 
Engineering Research Council of Canada through a subatomic 
physics grant. I very much enjoyed the hospitality
of the Kavli Institute for Cosmological Physics during 
my sabbatical, and I benefitted greatly from discussions with
KICP members and the particle cosmology group, especially
Rocky Kolb, Lian-Tao Wang, Andrew Long, Michael Fedderke,
and Dan Hooper. I also wish to acknowledge Valeri Galtsev for
his excellent computer support at KICP. 

\appendix
\section*{Appendix: The Schr\"odinger equation in the dark sector}
\setcounter{section}{1}

To understand the impact of the chiral projectors in the dark sector
Coulomb potential (\ref{eq:hrr1}), we follow the procedure which
yields the corresponding Schr\"odinger equation for the hydrogen
atom in the baryonic sector while keeping track of the chiral 
projectors.

The relevant states for the dark atoms are two-particle
states 
\begin{equation} \label{eq:2pstate1}
\bm{|}\Psi(t)\bm{\rangle}=\sum_{\alpha\beta}\int\!d^3\bm{x}\int\!d^3\bm{x}'\,
\Psi_{\alpha\beta}(\bm{x},\bm{x}',t)
\psi^+_{1\alpha}(\bm{x})\psi^+_{2\beta}(\bm{x}')\bm{|}0\bm{\rangle}.
\end{equation}
The states are written in the Schr\"odinger picture, and the indices
 $\alpha,\beta$ are Dirac indices.

The relevant part of the Hamiltonian of the theory in the sector of Fock
space wich is spanned by the states (\ref{eq:2pstate1}) (i.e. suppressing
all parts of the Hamiltonian which map into different sectors of Fock space)
is with $P_R=(1+\gamma_5)/2$,
\begin{eqnarray} \nonumber
H&=&\sum_{i,\alpha,\beta}\int\!d^3\bm{x}\,
\overline{\psi}_{i\alpha}(\bm{x})\left(
m_i\gamma_{\alpha\beta}-i\bm{\gamma}_{\alpha\beta}\cdot\bm{\nabla}\right)
\psi_{i\beta}(\bm{x})
\\ \label{eq:hrr2}
&&
-\sum_{\alpha,\beta,\rho,\sigma}P_{R\alpha\beta}P_{R\rho\sigma}
\int\!d^3\bm{x}\int\!d^3\bm{x}'\,\frac{\alpha_R}{|\bm{x}-\bm{x}'|}
\psi^+_{1\alpha}(\bm{x})\psi^+_{2\rho}(\bm{x}')
\psi_{2\sigma}(\bm{x}')\psi_{1\beta}(\bm{x}).
\end{eqnarray}

The corresponding Hamiltonian without the chiral projectors $P_R$
arises for the electron-proton system from the energy-momentum
tensor of QED in Coulomb gauge, see e.g.~Sec.~21.4 in Ref.~\cite{rdqm}.

The relativistic Schr\"odinger equation for the two-particle 
system follows from $id\bm{|}\Psi(t)\bm{\rangle}/dt=H\bm{|}\Psi(t)\bm{\rangle}$
and after decomposition in the basis (\ref{eq:2pstate1}) in the form
\begin{equation} \label{eq:SEQ1a}
i\frac{\partial}{\partial t}\Psi_{\alpha\beta}(\bm{x},\bm{x}',t)
=\sum_{\rho\sigma}\hat{H}_{\alpha\beta,\rho\sigma}\Psi_{\rho\sigma}(\bm{x},\bm{x}',t),
\end{equation}
with the Hamilton operator
\begin{eqnarray} \nonumber
\hat{H}_{\alpha\beta,\rho\sigma}&=&
\left[m_1\gamma^0_{\alpha\rho}-i(\gamma^0\cdot\bm{\gamma})_{\alpha\rho}\cdot
\frac{\partial}{\partial\bm{x}}\right]\delta_{\beta\sigma}
+\delta_{\alpha\rho}
\left[m_2\gamma^0_{\beta\sigma}-i(\gamma^0\cdot\bm{\gamma})_{\beta\sigma}\cdot
\frac{\partial}{\partial\bm{x}'}\right]
\\ \label{eq:SEQ1b}
&&-\frac{\alpha_R}{|\bm{x}-\bm{x}'|}P_{R\alpha\rho}P_{R\beta\sigma}.
\end{eqnarray}

To derive the nonrelativistic limit, we use the Dirac
representation of $\gamma$ matrices and write the $4\times 4$
matrix $\Psi_{\alpha\beta}(\bm{x},\bm{x}',t)$ in terms of $2\times 2$ 
matrices in the form
\begin{equation} \label{eq:nonr1}
\underline{\Psi}(\bm{x},\bm{x}',t)
=\left(\begin{array}{cc}
\underline{\psi}(\bm{x},\bm{x}',t) & \underline{\phi}(\bm{x},\bm{x}',t)\\
\underline{\xi}(\bm{x},\bm{x}',t) & \underline{\chi}(\bm{x},\bm{x}',t)\\
\end{array}\right)
\exp[-i(m_1+m_2)t].
\end{equation}
Substitution into Eq.~(\ref{eq:SEQ1a}) then yields in leading order
of $m_i^{-1}$ and $\alpha_R$ for the ``small'' components the equations
\[
\underline{\phi}(\bm{x},\bm{x}',t)=-\frac{i}{2m_2}
\frac{\partial}{\partial\bm{x}'}\underline{\psi}(\bm{x},\bm{x}',t)
\cdot\underline{\bm{\sigma}}^T,
\]
\[
\underline{\xi}(\bm{x},\bm{x}',t)=-\frac{i}{2m_1}\underline{\bm{\sigma}}\cdot
\frac{\partial}{\partial\bm{x}}\underline{\psi}(\bm{x},\bm{x}',t),
\] 
and $\underline{\chi}(\bm{x},\bm{x}',t)=0$, and substitution into 
the equation following for $\underline{\psi}(\bm{x},\bm{x}',t)$ from
Eq.~(\ref{eq:SEQ1a}),
\begin{equation} \label{eq:SEQ2}
i\frac{\partial}{\partial t}\underline{\psi}=
-i\underline{\bm{\sigma}}\cdot
\frac{\partial}{\partial\bm{x}}\underline{\xi}
-i\frac{\partial}{\partial\bm{x}'}\underline{\phi}
\cdot\underline{\bm{\sigma}}^T
-\frac{\alpha_R}{4|\bm{x}-\bm{x}'|}\left(
\underline{\psi}+\underline{\phi}+\underline{\xi}+\underline{\chi}
\right),
\end{equation}
yields the standard two-particle Schr\"odinger equation up to an
extra factor of $1/4$ in the Coulomb potential,
\begin{equation} \label{eq:SEQ3}
i\frac{\partial}{\partial t}\underline{\psi}(\bm{x},\bm{x}',t)=
\left(-\frac{1}{2m_1}\Delta-\frac{1}{2m_2}\Delta'\right)
\underline{\psi}(\bm{x},\bm{x}',t)
-\frac{\alpha_R}{4|\bm{x}-\bm{x}'|}
\underline{\psi}(\bm{x},\bm{x}',t).
\end{equation}
Separation in center of mass and relative coordinates for
the mapping into the effective single-particle equations with
masses $M=m_1+m_2$ and $m_{12}=m_1 m_2/M$ then proceeds as usual.

\end{document}